\documentstyle[aps,twocolumn,graphicx,floats,prl,amsmath]{revtex} 
\DeclareMathOperator{\trace}{Tr}
\newcommand{\pf}{{\cal Q}}
\newcommand{\order}[1]{{\cal O}\left(#1\right)}

\begin{document}
\wideabs{

\title{Weak Long-Ranged Casimir Attraction in Colloidal Crystals}
\author{Ajay Gopinathan, Tong Zhou\footnotemark, S. N. Coppersmith\footnotemark, L. P. Kadanoff, 
  and David G. Grier}

\address{Dept. of Physics and James Franck Institute,
The University of Chicago, 5640 S. Ellis Avenue,  Chicago, IL 60637}

\date{\today}

\maketitle

\begin{abstract}
We investigate the influence of geometric confinement
on the free energy of an idealized model for charge-stabilized
colloidal suspensions.
The mean-field 
Poisson-Boltzmann formulation for this system predicts pure repulsion
among macroionic colloidal spheres.
Fluctuations in the simple ions' distribution
provide a mechanism
for the macroions to attract each other at large separations.
Although this Casimir interaction is long-ranged, it is
too weak to influence colloidal crystals' dynamics.
\end{abstract}
% 82.70.Dd Colloids
% 05.20.Jj Statistical mechanics of classical fluids
% 24.10.Pa Thermal and statistical models
% 05.70.Ce Thermodynamic functions and equations of state
% 61.20.Qg Structure of associated liquids: electrolytes, molten salts, etc.
} % end of wideabs

Experimental evidence collected over 20 years \cite{ise96a}
suggests that similarly-charged
colloidal spheres dispersed in water need not simply repel each other.
Under some circumstances they instead experience an
unexpected long-ranged attraction.
For example, like-charge attractions are implicated in the
cohesion of metastable superheated colloidal crystals \cite{larsen96,larsen97}
even though isolated pairs of the constituent
spheres are observed to repel each other
\cite{repel,crocker96a}.
Comparable attractions have been measured for pairs of spheres confined by
 two 
\cite{crocker96a,attract} charged planar walls.
Recent calculations \cite{neu99,Trizac}
reveal that such confinement-induced
attractions cannot be accounted
for by local density theory nor by electrohydrodynamic coupling \cite{squires,grier2}.
Such anomalous effects in charge-stabilized colloid
therefore challenge our general
understanding of interactions and dynamics in macroionic systems.

This Letter addresses fluctuations'
contribution to the free energy of highly
charged colloidal spheres surrounded by a neutralizing
cloud of small singly-charged counterions.
Highly symmetric
monopolar fluctuations in the counterion distribution
increase the system's free energy.
We demonstrate that
their suppression by boundary conditions at the spheres'
surfaces introduces a long-range attraction into the
crystal's free energy analogous to the Casimir force 
in quantum electrodynamics, but that it is too weak to
account for anomalous behavior in charge-stabilized
suspensions.\footnotetext{$\ast$ Present address :Physics Department, University of California, Santa Barbara, CA 93106-9530} \footnotetext{$\dag$ Present address :Physics Department, University of Wisconsin, 1150 University Avenue, Madison, WI 53705}

Our treatment is based on
the Wigner-Seitz cell model
which has been studied extensively \cite{attard96}
both theoretically and through Monte Carlo simulation.
It consists of a single spherical macroion of radius $a$ carrying a
uniformly distributed surface charge $-Ze$ and surrounded by a thermal
cloud of $Z$ point-like counterions at temperature $T$, 
each carrying a single charge $e$.
The macroion and counterions are confined by a concentric
conducting spherical shell of radius $R$.
This outer shell plays a role
analogous to the Wigner-Seitz
cell boundary in a colloidal crystal.
More generally, it models the crowding or geometric confinement
characteristic of colloidal crystals \cite{grier00}.

Previous investigations of this and related models 
\cite{attard96,kardar99}
have found
short-ranged correlation-driven attractions between the bounding
surfaces under some 
conditions,
particularly when the  counterions are polyvalent.
They have not found evidence for 
long-ranged attractions in monovalent electrolytes
\cite{grier00}.

Our method for evaluating the counterions' partition function
allows us to investigate much higher macroion charges than 
have been considered before.
The outer boundary's suppression of counterion fluctuations
induces a long-ranged Casimir-like attraction \cite{kardar99}
between the macroion and its neighbors across the
Wigner-Seitz cell boundary.
Although this cell model is far too simple to describe the behavior
of real charge-stabilized suspensions, it highlights a previously
unexplored mechanism for long-ranged confinement-induced
like-charge colloidal attractions.

We adopt the path integral formalism reviewed in \cite{kardar99}
and write the counterions' canonical partition
function as a functional integral
over all possible counterion distributions,
$n({\vec r})$:
\begin{equation}
  \label{eq:partition}
  \pf = \sqrt{2\pi Z} \, Z^Z \, \int^\prime e^{-\beta f[n]} \, Dn,
\end{equation}
where $\beta^{-1} = k_B T$ is the thermal energy scale at temperature $T$, 
the prime indicates that the number of charges is conserved
($\int n \, dV = Z$),
and  \cite{zhou00}
\begin{equation}
  \label{eq:weight}
  f[n] = U[n] + k_B T \, \int n \, \ln n \, dV.
\end{equation}
The potential energy functional
\begin{equation}
  \label{eq:potential}
  U[n] = \frac{1}{2} \int ne \phi \, dV,
\end{equation}
describes the counterions' interaction with the local
electric potential $\phi({\vec r})$.
The system's Helmholtz free energy is then $F = -k_B T \ln \pf$.

One ionic distribution, $n_0({\vec r})$, minimizes $f$,
and thus has the greatest statistical
weight in $\pf$.
We factor the partition function
$\pf = \pf_0 \pf_{fl}$
into the saddle point contribution
\begin{equation}
  \pf_0 = Z^Z \, e^{-\beta f_0},
  \label{eq:saddle}
\end{equation}
where $f_0 = f[n_0]$,
and a term $\pf_{fl}$ accounting for fluctuations, $\delta n$,
away from $n_0$.
Expressing $\pf_{fl}$ as a series expansion in $\delta n$ yields
as the lowest-order non-vanishing term
\begin{equation}
  \pf_{fl} \simeq \sqrt{2\pi Z} \, \int^\prime \, e^{-\beta \delta^2f} \; Dn,
  \label{eq:master}
\end{equation}
where $\delta^2f$ is the second-order change in $f[n]$ due
to $\delta n$. 

Terminating this expansion at Gaussian order is justified
if corrections at higher order in $\delta n$ contribute
negligibly to $\pf_{fl}$.
This condition is met if $Z$ is large \cite{Anchang} and $n_0(r)$ itself changes negligibly over
the mean radial counterion separation:
\begin{equation}
  \frac{1}{n_0} \, \frac{dn_0}{dr} \ll 4 \pi r^2 n_0,
  \label{eq:condition}
\end{equation}
and can be tested \emph{a posteriori} once $n_0(r)$
is evaluated.
Expanding around any other distribution 
\cite{expansions,netz2} would not give such a
criterion for establishing convergence. 

It has been shown \cite{Anchang,netz2} that the saddle point corresponds
to the mean-field result
\begin{equation}
  n_0 = n_s e^{-\beta e \phi}
\end{equation}
which, combined with the Poisson equation
\begin{equation}
  \nabla^2 \phi = - \frac{en}{\epsilon},
  \label{eq:poisson}
\end{equation}
yields the familiar Poisson-Boltzmann (PB) equation,
\begin{equation}
  \nabla^2 \phi = - \frac{e n_s}{\epsilon} \, e^{-\beta e \phi}.
  \label{eq:pb}
\end{equation}
Selecting
$n_s = 3Z/(4\pi a^3)$ sets the potential's reference point
conveniently without loss of generality. 
Eq.~(\ref{eq:poisson}) accounts for the solvent's influence
in the so-called primitive
model through its dielectric constant $\epsilon$.

\begin{figure}
  \centering
  \includegraphics[width=3.2in]{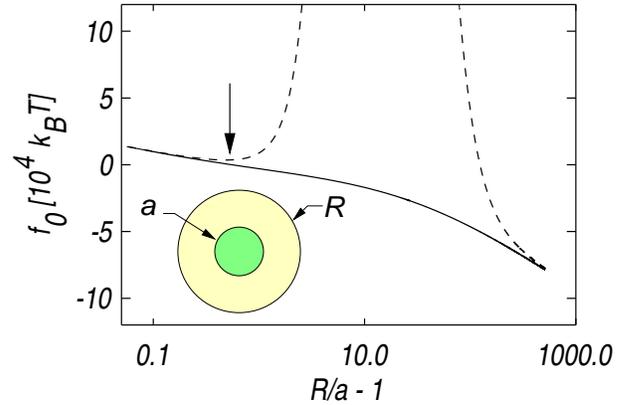}
  \vspace{1ex}
  \caption{Dependence of the mean-field free energy on the confining
    boundary's radius, $R$, from Eq.(14) with $Z=7300$,
    $a = 0.326 \mu$m and
    $\lambda_B = 0.7 $ nm. 
    The solid curve shows the solution of the full
    PB equation while the dashed line is from the linearized PB theory.}
  \label{fig:f0}
\end{figure}

Following conventional practice, we introduce the Bjerrum length
$\lambda_B = \beta e^2 / (4 \pi \epsilon)$
and an effective screening
length $\kappa^{-1} = (4 \pi n_s \lambda_B)^{-1/2}$.
Taking the
system's radial symmetry into account leads to
\begin{equation}
  \frac{d^2\phi}{dr^2}+\frac{2}{r}\frac{d\phi}{dr}
  - \frac{\kappa^2}{e \beta} e^{- \beta e \phi}=0.
  \label{eq:nondi}
\end{equation}
We solve Eq.~(\ref{eq:nondi}) subject to two boundary
conditions:
Gauss' theorem at the macroion's surface gives
\begin{equation}
  \left. \frac{d\phi}{dr} \right|_{r= a} = \frac{\kappa^2 a}{3\beta e},
\end{equation}
and electroneutrality requires
\begin{equation}
  \left. \frac{d\phi}{dr} \right|_{r = R} = 0.
  \label{eq:neutral}
\end{equation}
Equations (\ref{eq:nondi}--\ref{eq:neutral}) can be solved numerically
for $\phi(r)$ from which we can calculate the mean-field free energy,
\begin{align}
  f_0 & = - \frac{Z}{2} \, \left[
    e \phi(a) + 
    \frac{3e}{a^3} \, \int_a^R
    \phi \, e^{-\beta e \phi} \, r^2 \, dr \right] \\
  & = \frac{Z}{2} \, k_B T \, \left[
    \ln \left( \frac{n_0(a)}{n_s} \right) + 
    \frac{3}{a^3} \, \int_a^R
    \frac{n_0}{n_s} \, \ln \left(\frac{n_0}{n_s}\right) \, r^2 dr \right].
  \label{eq:f0}
\end{align}
Figure~\ref{fig:f0} shows $f_0$ as a function of $R/a$, calculated 
numerically from the solution of the full PB equation.  
Since $f_0(R)$ decreases monotonically with $R$, the mean-field theory
predicts pure repulsion.

An equivalent result obtained by solving the linearized PB
equation for the same system \emph{does} have a local minimum
for $R$ slightly bigger than $a$. 
In the context of colloidal interactions, the linearized
mean-field description yields the electrostatic component of
the pair potential due to Derjaguin, Verwey, Landau
and Overbeek (DLVO) \cite{dlvo}.
The local minimum in the free energy suggests an effective
electrostatic attraction between like-charged colloidal
spheres within the DLVO theory.
Comparison with the full calculation, however,
shows this to be an artifact of linearization.

In addition to confirming the absence of attractions in
the mean-field description, our numerical results also
satisfy the condition in Eq.~(\ref{eq:condition}).
Thus, we are justified in using Eq.~(\ref{eq:master}) to calculate $\pf_{fl}$.
The second-order change in $U[n]$ due to $\delta n$ is
\begin{equation}
  \delta^2U = \frac{e^2}{8\pi\epsilon } \, 
  \int d^3r_1\int d^3r_2 \,
  \frac{\delta n(\vec{r}_1)\delta n(\vec{r}_2)}{|\vec{r}_1-\vec{r}_2|}.
  \label{eq:fluctuation}
\end{equation}
So, from Eq.~(\ref{eq:master}),
\begin{equation}
  \pf_{fl} = \sqrt{2\pi Z} \, 
  \int^\prime  
  \exp \left[-\beta\delta^2U-\int\frac{(\delta n)^2}{2n_0} dV \right] \, Dn.
\end{equation}
We evaluate the functional integral by partitioning the system
into $N$ concentric shells,
\begin{multline}
  \pf_{fl} = \sqrt{2\pi Z} \,
  \int^\prime \prod_{i=1}^N \frac{d\delta n_i}{\sqrt{2\pi n_{0i}/V_i}} \\
  \exp \left[
    -\sum_{j=1}^N \frac{(\delta n_j)^2}{2n_{0j}/V_j}
     - \frac{\lambda_B }{2} \sum_{j,k=1}^N \delta n_j\delta n_k
    \int_{V_j} \int_{V_k} \frac{d^3r_jd^3r_k}{|\vec{r}_j-\vec{r}_k|}
  \right],
\end{multline}
where $V_i$ is the volume of cell $i$, $n_i$ the number density 
of counterions therein,
and $n_{0i}$ the equivalent number density for the mean-field case.
This highly symmetric partition of the system is appropriate for
radial or monopolar density fluctuations.
A more complete description including multipole fluctuations
is not necessary for our purposes because recent calculations
reveal that these yield only
short-ranged attractions \cite{levin99}.
 Cross terms cancel at the Gaussian level of approximation,
so we can consider monopole and multipole contributions
separately.

We define $x_i = \delta n_i/\sqrt{2n_{0i}/V_i}$, and divide
the system so that every cell has the same number of counterions in
the mean-field distribution, \emph{i.e.} $n_{0i}V_i = Z/N$.
Rescaled in this way,
\begin{align}
  \pf_{fl} & = \sqrt{2\pi Z} \, 
  \int^\prime \prod_{i=1}^N \frac{dx_i}{\sqrt{\pi}} \label{eq:general}\\
  & \exp \left[ - \sum_{j=1}^N {x_j}^2 
  - \frac{Z\lambda_B}{N} \sum_{j,k=1}^N
    \frac{x_j x_k}{V_j V_k} \int_{V_j} \int_{V_k} 
    \frac{d^3r_j d^3r_k}{\left| \vec r_j - \vec r_k \right|} \right]. \nonumber
\end{align}
Then, numbering the cells from the center outwards
so that $r_i < r_j$ if $i<j$, we obtain
\begin{align}
  \pf_{fl} &= \sqrt{2\pi Z}\int^\prime \prod_{i=1}^N 
  \frac{dx_i}{\sqrt{\pi}} \\
  & \exp\left[
    -\sum_{j=1}^N\left(1+\frac{Z\lambda_B}{N}
      \frac{1}{r_j}\right)x_j^2
    -\frac{Z \lambda_B}{N}\sum_{j<k}
    \frac{2}{r_k}x_jx_k
  \right] \nonumber\\
  & = \sqrt{2\pi Z}\int^\prime \prod_{i=1}^N\frac{dx_i}{\sqrt{\pi}}
  \exp\left[-\vec{x}\cdot {\mathbf{A}} \cdot\vec{x}\right].
\end{align}
The prime indicates that the integral is constrained to maintain
electroneutrality: $\sum_{i=1}^N x_i=0$.  Because of this condition,
we can subtract a constant from each element of $\mathbf{A}$ without changing
the value of ${\mathbf{A}}\cdot\vec{x}$.  
Let us subtract 
$Z \lambda_B/(N R)$ from each element of
$\mathbf{A}$ to obtain matrix $\mathbf{\tilde A}$.
All of the elements in the
last row and column of $\mathbf{\tilde A}$ vanish except ${\mathbf{\tilde A}}_{N,N}$ which is $1$.
Separating mode $x_N$ from the other $N-1$ modes in this way
allows us to remove the explicit constraint:
\begin{gather}
  \int^\prime \prod_{i=1}^N\frac{dx_i}{\sqrt{\pi}} \, e^{-{x_N}^2}
  = \int\prod_{i=1}^N \frac{dx_i}{\sqrt{\pi}} \,
  e^{-{x_N}^2} \,  
  \delta\left(\sqrt{\frac{2Z}{N}} \sum_{j=1}^N x_j\right) \\
  = \sqrt{\frac{N}{2\pi Z}} \, 
  \int \prod_{i=1}^{N-1}\frac{dx_i}{\sqrt{\pi}} \;
  \exp\left[-\left(\sum_{j=1}^{N-1}x_j\right)^2\right].  
\end{gather}
Consequently,
\begin{equation}
  \pf_{fl} = \sqrt{N} \, \int
  \prod_{i=1}^{N-1}\frac{dx_i}{\sqrt{\pi}} \;
  e^{-\vec{x}\cdot {\mathbf{B}} \cdot\vec{x}}
  = \sqrt{\frac{N}{\det {\mathbf{B}} }}, \label{eq:zm}
\end{equation}
where $\vec{x}$ is now a vector of dimension $N-1$, and ${\mathbf{B}}$ is
a matrix of dimension $N-1$ obtained by adding $1$ to each element
in the first $N-1$ rows and columns of matrix ${\mathbf{\tilde A}}$.

$\mathbf{B}$ may be expressed as the sum of two matrices, $\mathbf{C}$ and $\mathbf{D}$, 
whose components are
${\mathbf{C}}_{ij} = 1 + \delta_{ij}$ and
\begin{equation}
 {\mathbf{D}}_{ij} = \frac{Z}{N} \, \lambda_B \, \left(
    \frac{1}{r_p} - \frac{1}{R} \right),
\end{equation}
where $p$ is the greater of $i$ and $j$.
$\det {\mathbf{C}} = N$, so that
\begin{equation}
 \det {\mathbf{B}} = N \, \det {{\mathbf{I}} + {\mathbf{C}}^{-1}{\mathbf{D}}},
\end{equation}
where $\mathbf{I}$ is the identity matrix.

Evaluating $\det \mathbf{B}$ is greatly facilitated
if the components of ${\mathbf{C}}^{-1}{\mathbf{D}}$ are much smaller than 1.
${\mathbf{C}}_{ij}^{-1} = \delta_{ij} - 1/N$ differs little from the
identity matrix.
The components of $\mathbf{D}$, on the other hand, are bounded above by
${\mathbf{D}}_{ij} < (Z/N) \, (\lambda_B/a)$.
We previously assumed  $Z/N \gg 1$ in deriving
Eqs.~(\ref{eq:saddle}) and (\ref{eq:master}).
But $\lambda_B/a \ll 1$ for the micron-sized spheres
in experimental observations, so that we may
reasonably assume ${\mathbf{D}}_{ij} < 1$.
Even if this were not the case, we would be justified in
formally taking the limit $Z/N \ll 1$
at this point because the final result cannot depend on $N$.
Consequently,
\begin{align}
  \det {\mathbf{B}}  & = N \, \exp\left[ \trace \ln \left({\mathbf{I}} + {\mathbf{C}}^{-1}{\mathbf{D}}\right) \right] \\ & \approx N \, \exp \left( \trace {\mathbf{C}}^{-1}{\mathbf{D}}\right).
 \end{align}
In this approximation, the fluctuation contribution to the free energy is
\begin{align}
  \delta F & = \frac{1}{2} \, k_B T \, \trace {\mathbf{C}}^{-1}{\mathbf{D}} \\
  & = \frac{Z}{2} \, k_B T \, \frac{1}{N^2} \, \sum_{k=1}^{N-1} \sum_{j=1}^k \left( \frac{\lambda_B}{r_j} - \frac{\lambda_B}{r_{k+1}} \right).
\end{align}
Rewriting the sums over shell indices as integrals over radii, we
obtain
\begin{multline}
  \delta F \approx \frac{Z}{2} \, k_B T \, 
  \left( \frac{4\pi}{Z} \right)^2 \,
  \int_a^R dr \, r^2 \, n_0(r) \times \\
  \int_a^r dr^\prime \, {r^\prime}^2 n_0(r^\prime) \,
  \left( \frac{\lambda_B}{r^\prime} - \frac{\lambda_B}{r} \right).
\end{multline}

Most of the boundaries, $r_i$, between cells in the
mean-field distribution
are clustered near $a$.
Consequently,
\begin{gather}
  \delta F \approx   2 \pi k_B T \,
  \int_a^R  n_0(r) \, w_0(r) \, 
  \left( \frac{\lambda_B}{a} - \frac{\lambda_B}{r} \right) \, r^2 dr,
\end{gather}
where 
$w_0(r) = (4\pi/Z) \, \int_a^r n_0(r^\prime) \, {r^\prime}^2 \, dr^\prime$ 
is the
fraction of counterions within radius $r$, in the mean field approximation.
Integrating by parts then yields
\begin{equation}
  \delta F \approx \frac{Z}{2} \, k_B T \, \lambda_B \,
  \int_a^R \frac{1 - w_0^2(r)}{r^2} \, dr.
 \label{eq:ffl}
\end{equation}

\begin{figure}
  \centering
  \includegraphics[width=3.2in]{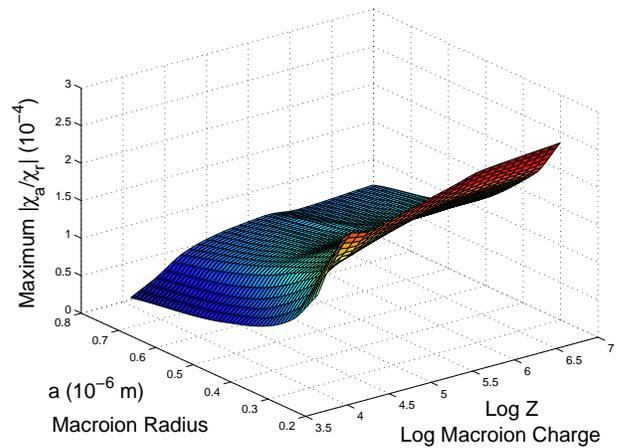}
  \vspace{1ex}
  \caption{Numerically evaluated ratio $\vert \chi_a/\chi_r\vert$
    maximised over $R$ for various values of the macroion radius $a$ and
    the macroion charge $Z$. A smooth surface is fitted to the data points
    to aid the eye. The plot clearly demonstrates that the attraction is
    four orders of magnitude too small to overcome the repulsion.}
  \label{fig:f1}
\end{figure}

Unlike $f_0(R)$, $\delta F(R)$ decreases with decreasing $R$
because the outer boundary condition suppresses
fluctuations as $R$ approaches $a$.
The resulting attraction therefore is reminiscent of
the Casimir attractions previously identified in confined electrolytes
as well as other systems \cite{kardar99}. It is interesting to note that 
monopolar fluctuations do not yield an attractive contribution in all geometries;
the second term in the exponent of Eq.~(\ref{eq:general}) vanishes
for unbounded systems such as parallel plates and concentric
cylinders. This is consistent with the absence of long-ranged like-charge
attractions in measurements \cite{israelachvili92}, theoretical treatments
\cite{attard96,kardar99}, and
simulations \cite{attard96,kardar99} of unbounded systems.
Such attractions, therefore, are peculiar to closed systems, such
as the Wigner-Seitz cells of colloidal crystals.
If there is an $R$ at which the  attractive force
\begin{equation}
  \chi_a \equiv - \frac {\partial \delta F(R)}{\partial R} =
  \frac{Zk_{B}T \lambda_{B}}{2}
  \int_a^R \frac{2\omega_{0}(r)}{r^2} 
  \frac{\partial \omega_{0}(r)}{\partial R} dr
\label{eq:flderiv}
\end{equation}
has larger absolute value than the mean-field repulsive force
\begin{equation}
\begin{split}
  \chi_r  & \equiv -\frac{\partial f_0(R)}{\partial R} = 
  -\frac{Zk_{B}T}{2} \left[ 
    n_{0}(R) \ln \left(\frac{n_{0}(R)}{n_{s}}\right ) R^{2} \right. \\
  & \left. + \int_{a}^{R} \left ( \frac{\partial n_{0}(r)}{\partial R} 
      + \frac{\partial n_{0}(r)}{\partial R} 
      \ln \left(\frac{n_{0}(r)}{n_{s}}\right ) \right ) r^2 dr \right],
\end{split}
\end{equation}
then
$F(R) = f_0(R) + \delta F(R)$ would have a minimum at that $R$.
Such a minimum would correspond to a fluctuation-mediated 
bound state for a colloidal crystal of nearest neighbor spacing $2R$.
However, numerical results shown in Fig.~\ref{fig:f1} reveal
$\vert \chi_a/\chi_r\vert < 10^{-4}$ over the entire 
range of conditions studied
experimentally.
On this basis, we conclude that monopolar
fluctuations are not responsible for the strong and long-ranged
attractions reported in measurements on charge-stabilized colloid
\cite{larsen96,larsen97,crocker96a,attract}.

To understand why the ratio $\vert \chi_a/\chi_r\vert$ is small we estimate it  analytically by assuming that most of the counterions are clustered
close to the macroion surface and that only a {\it small} fraction
$p(Z,a,R)$ of counterions are affected by a change in the radius $R$
of the confining shell, and that this fraction is uniformly
distributed in the volume.  
Evaluating $\chi_a(R)$ to $\order{p}$
then yields
\begin{equation}
  \vert \chi_a \vert = 
  \frac{Zk_{B}T \lambda_{B}}{2} \frac{3p}{2} \frac{x(x+2)}{a^2(x^2 +1 + x)^2}
\end{equation}
where $ x = R/a $.
Similarly evaluating $\chi_r(R)$ yields 
\begin{equation}
 \vert \chi_r  \vert = \frac{Zk_{B}T }{2} \frac{3 p }{2a}\frac{x^2}{x^3-1}
\end{equation}
Truncating to $\order{p}$ is justified by numerical 
investigation of the mean-field solution which indicates 
$p \leq 0.05$ in the region of interest. 
In this approximation the ratio 
\begin{equation}
  \frac{\vert \chi_a \vert}{\vert \chi_r  \vert} = 
  \frac{\lambda_B}{a} \frac{(x^2+x-2)}{(x^2+x+1)} \frac {1}{x} 
\end{equation}
is independent of $p$ and has a maximium value $0.286~\lambda_B/a$ at
$R = 2.067 a$.
The location and magnitude of the maximum value agrees with
the full numerical solution to within factors of 1.2 and 4, respectively.
This value of $R$ would correspond to half the nearest-neighbor
separation in a fluctuation-stabilized colloidal crystal; it
also is consistent with the lattice constants observed in superheated
metastable colloidal crystallites \cite{larsen96,larsen97}.
However, water has a dielectric constant $\epsilon = 80$ which
corresponds to a Bjerrum length $\lambda_B = 0.7$ nm at
room temperature.
Consequently, $| \chi_a / \chi_r | \ll 1$ for all reasonable
colloidal radii.
Changing solvents would not affect this conclusion because of the
limited range of accessible values of $\epsilon$.

We also investigated the possible influence of 
multivalent counterions carrying charge $qe$. 
Ignoring the relatively weak $Z$ dependence of
$\chi_a/\chi_r$ yields $\chi_a/\chi_r \sim q^2$. 
Despite the relative
strength of the attraction being larger for higher valency
counterions, the effect is still too weak to induce measurable
attractions for physically plausible values of $q$.
 
We have demonstrated that suppression of monopolar
ionic fluctuations by
surfaces induces a long ranged attraction remniscent of Casimir
attractions \cite{kardar99}. 
This interaction is
distinct from and complementary to 
attractions driven by multipolar fluctuations which have been
studied elsewhere \cite{levin99}.
Neither mechanism, however,
accounts for the strong and long ranged attractions observed
experimentally between highly charged colloidal spheres.

The long-ranged like-charge attractions observed in confined colloid
are not consistent with mean field theories for electrolyte
structure.
Possible explanations must incorporate mechanisms such as
fluctuations and high-order correlations not captured by mean field
theory.
While multipole fluctuations in the distribution of simple ions
induce strong attractions \cite{levin99}, they are short ranged.
The present study demonstrates that radially symmetric fluctuations can
induce long-ranged attractions, but that they are too weak to 
influence colloidal behavior.
Consequently, the explanation must lie in another mechanism not
yet considered and thus remains an important outstanding challenge.

We are grateful to Tom Witten, Stuart Rice, Phil Pincus, 
Mehran Kardar, Ramin Golestanian, An-Chang Shi and Adrian Parsegian
for enlightening conversations.  This work was supported by the
MRSEC Program of the National Science Foundation under Award Number
DMR-9808595.

\end{document}